\documentclass{ws-ijmpd}
\usepackage[super,compress]{cite}

\begin{document}

\markboth{N. Sahakyan}
{AI in the Cosmos}

%
\catchline{}{}{}{}{}
%

\title{AI in the Cosmos
}

\author{N. Sahakyan
}

\address{ICRANet-Armenia, Marshall Baghramian Avenue 24a, Yerevan 0019, Armenia
\\
narek.sahakyan@icranet.org}



\maketitle

\begin{history}
\received{Day Month Year}
\revised{Day Month Year}
\end{history}

\begin{abstract}
Artificial intelligence (AI) is revolutionizing research by enabling the efficient analysis of large datasets and the discovery of hidden patterns.  In astrophysics, AI has become essential, transforming the classification of celestial sources, data modeling, and the interpretation of observations. In this review, I highlight examples of AI applications in astrophysics, including source classification, spectral energy distribution modeling, and discuss the advancements achievable through generative AI. However, the use of AI introduces challenges, including biases, errors, and the "black box" nature of AI models, which must be resolved before their application. These issues can be addressed through the concept of Human-Guided AI (HG-AI), which integrates human expertise and domain-specific knowledge into AI applications. This approach aims to ensure that AI is applied in a robust, interpretable, and ethical manner, leading to deeper insights and fostering scientific excellence.

\end{abstract}


\ccode{PACS numbers:}


\section{Introduction}
Astronomy and astrophysics have always been, and will remain, data-driven sciences, relying heavily on experimental measurements to develop theories about the universe and the processes occurring in various sources. Early astronomical catalogues included the coordinates of various objects visible to the naked eye, recorded over different periods which was limited, but it formed the fundamental basis for any theories or hypotheses that attempted to explain the bright phenomena observed in the night sky. With advancements in telescopes—particularly when all-sky surveys are available (for example, Sloan Digital Sky Survey \cite{2000AJ....120.1579Y} and the upcoming Vera C. Rubin Observatory \footnote{https://rubinobservatory.org}) - the volume of astronomical data now reaches terabyte and petabyte scales. This immense quantity of data has provided researchers with a unique opportunity to study the cosmos with unprecedented detail.

As the volume of astronomical data has increased, the methods for analyzing and interpreting it have evolved accordingly. In the 1970s and 1980s, when datasets were smaller and the number of observed sources was manageable, traditional statistical methods were sufficient for data exploration. Today, however, far more sophisticated methods, advanced comprehensive tools and algorithms are required to automate data processing and analysis, as the quantity of observed data significantly increased and because of complexity of observed sources. However, the continuous exponential growth of astronomical data presents a new challenge: while current tools and methods enable extraction of information, they are not enough to extract maximum information contained in the observed data. This limitation arises because identifying patterns or relationships within large datasets, a significant computational resources and time is required. In this context, machine intelligence - machine learning and artificial intelligence- can play a transformative role in advancing research. These methods can significantly advance research and overcome current limitations by efficiently handling vast amounts of data and identifying hidden patterns. 

To overcome the limitations of traditional data analysis methods, it is crucial to explore the potential of artificial intelligence (AI), machine learning (ML), and generative AI in processing astronomical data. AI refers to tools and systems that can perform tasks requiring human-like intelligence, such as recognizing patterns, and making decisions. ML, a subset of AI, uses algorithms and statistical models to allow computers to learn from data and improve their performance over time without being explicitly programmed for specific tasks. Generative AI is a specialized area within AI and ML that can understanding natural language, generate high-quality text, images, and other content based on the data on which they were trained. Incorporating AI, ML, and generative AI into data analysis can transform machines—traditionally smart systems capable of performing complex calculations at high speeds but lacking the ability to adapt to new information or learn from data—into adaptive learners, to become capable of identifying intricate patterns, generating new data, and uncovering relationships within large and complex datasets. This advancement is particularly necessary in astronomy, where the ability to efficiently process, analyze, and even simulate vast amounts of data can lead to significant discoveries and a deeper understanding of the universe.

In this era when astrophysical research is being transformed through the application of AI and ML methods, in this paper I discuss their recent applications in astrophysical studies. I discuss the future prospects with the arrival of generative AI, as well as the issues and challenges associated with the application of AI in research. Furthermore, I will argue that careful oversight application of AI in research, namely the concept of Human Guided Artificial Intelligence (HG-AI), can significantly contribute to the effectiveness of scientific investigations.

\section{ML used in astrophysical research}
\begin{figure}
    \centering
    \includegraphics[width=1\linewidth]{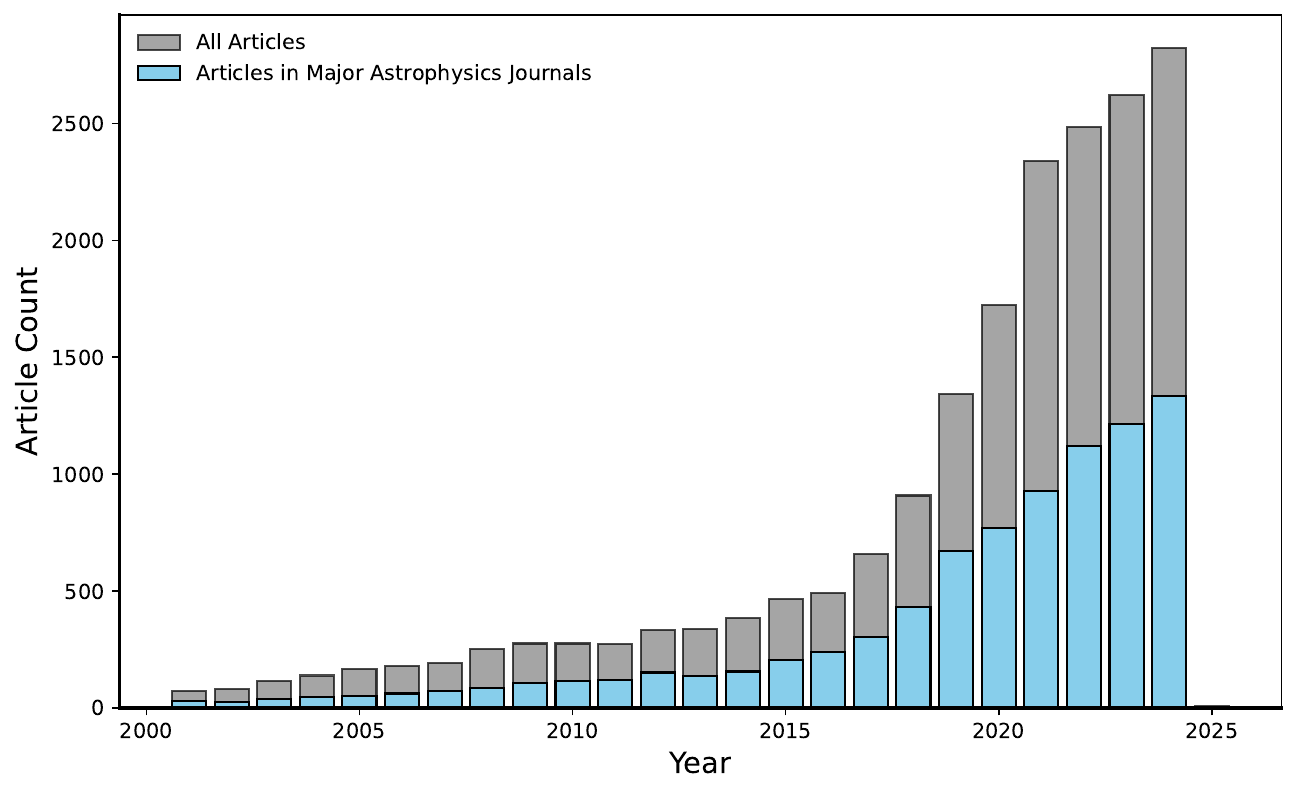}
    \caption{Number of astrophysics articles containing the term "machine learning" over the years. }
    \label{fig:art}
\end{figure}

The application of ML in astrophysics has grown remarkably in recent years. Fig. \ref{fig:art} shows the number of articles from NASA Astrophysics Data System \footnote{https://ui.adsabs.harvard.edu} containing the term "machine learning" in different years. The article number is shown by two categories: all articles (in gray) and those published specifically in major astrophysics journals (such as ApJ, A\&A, MNRAS, etc.) (light blue). The general trajectory reveals a clear acceleration in the adoption of ML methodologies within the astrophysics community. From around 2000 to 2010, the number of articles is low, indicating that ML was not yet widely adopted in astrophysics. However, starting in the early 2010s, there is a noticeable increase, with a rapid and nearly exponential growth beginning around 2015. This growth pattern suggests a transformative phase in astrophysical research, where different tools of ML were started to be used in data analysis and modeling works. The upward trajectory in Fig. \ref{fig:art} indicates that the role of ML methods in solving complex astrophysical problems, from analyzing large datasets to modeling, continues to grow. This highlights the recognition by the community of the potential of ML methods for data analysis and interpretation in the field.
\section{Application of ML for specific tasks}
Astrophysical data is increasing at an unprecedented rate due to advancements in observational technologies and large-scale surveys, making it impossible for humans alone to effectively handle and interpret the observed data. In this context, AI and ML tools have become essential for analyzing these large datasets by enabling algorithms to learn from data patterns and make predictions or decisions without explicit task-specific programming. These tools aim to solve problems in a manner analogous to human reasoning. The output of ML models includes classifications, predictions, clustering of data points, and anomaly detection, all of which are essential for extracting meaningful information from complex datasets. There are generally two main types of ML: supervised and unsupervised learning. Supervised learning involves training models using labeled data, where the expected output is known during the training process. This approach is particularly useful for tasks such as classifying objects, predicting different properties, and identifying specific astrophysical events. In contrast, unsupervised learning is used on unlabeled data, aiming to identify unknown structures or patterns. This method is used to tasks such as grouping galaxies based on their features, detecting new types of astronomical phenomena, and other similar applications.

Both supervised and unsupervised learning methods have been applied in astrophysical research for a range of tasks. A few examples include:

\begin{itemlist}

\item A random forest classifier was developed and trained on large samples of photometric and spectroscopic data from SDSS and LAMOST, achieving high accuracy in classifying stars, galaxies, and QSOs, enabling efficient large-scale classification of astronomical objects across multiple surveys \cite{2019AJ....157....9B}.

\item A convolutional neural network (CNN) trained on simulated images of galaxies at different evolutionary stages successfully retrieved various phases of evolution. This approach was subsequently applied to observed data, enabling the accurate distinction between pre-BN, BN, and post-BN stages \cite{2018ApJ...858..114H}.

\item CNN was trained on SDSS \textit{ugriz} images to estimate photometric redshifts for galaxies, using all pixel information, achieving high precision and reliability in redshift prediction across large datasets \cite{2019A&A...621A..26P}.

\item A deep neural network trained on labeled Kepler data effectively distinguished genuine transiting exoplanets from false positives by directly learning features from light curves, achieving high accuracy in identifying planet candidates \cite{2018AJ....155...94S}.

\item A deep learning framework trained on simulated gravitational wave templates enabled real-time detection and parameter estimation of gravitational waves directly from LIGO data \cite{2018PhLB..778...64G}.

\item  A gradient boosting decision tree model was trained on the spectral and temporal properties of BL Lacs and FSRQs from the Fermi Large Area Telescope (LAT) catalog, achieving high-accuracy classification of blazar candidates of uncertain type by learning distinct $\gamma$-ray emission patterns \cite{2023MNRAS.519.3000S}.

\item A machine learning tool trained on a large sample of blazar spectral energy distributions automated the estimation of synchrotron peak frequencies, allowing predictions of blazar synchrotron emission peaks even with limited observational data \cite{2022A&C....4100646G}.

\item Deep learning methods were developed to improve event selection, enabling accurate distinction of astrophysical neutrinos from background noise. This advancement enhanced the angular resolution and sensitivity of IceCube, facilitating the identification of neutrino emission from the Galactic plane \cite{2023Sci...380.1338I}.

\item CNN was trained on leptonic models that account for the injection of electrons, their self-consistent cooling, and pair creation-annihilation processes, considering both internal and external photon fields, which enable modeling the blazar emission with high accuracy \cite{2024ApJ...963...71B, 2024ApJ...971...70S}.

\end{itemlist}

The specific studies highlighted above, along with others, represent effective applications of AI/ML tools in research, helping to manage large datasets and sophisticated pattern-recognition challenges. These applications have not only increased our knowledge but have also maximized the usability of observed datasets by enabling the extraction of the maximum information contained within them. Below, I discuss two examples of ML applications in astrophysics: one related to a classification problem and the other to more complex modeling efforts. 
\subsection{Classification of blazars using high energy $\gamma$-ray data}
In Ref.~\refcite{2023MNRAS.519.3000S}, ML methods were applied to classify blazar candidates of uncertain type (BCUs) included in the Fermi-LAT 4FGL-DR3 catalogue \cite{2022ApJS..260...53A}. Blazars, which are active galactic nuclei with jets oriented toward Earth, are the dominant sources in the extragalactic $\gamma$-ray sky, as demonstrated by Fermi-LAT observations. They are broadly classified into two types: BL Lacertae objects (BL Lacs) and Flat Spectrum Radio Quasars (FSRQs), which are distinguished by their emission-line features. However, many $\gamma$-ray detected blazars lack clear classification due to limited optical spectroscopic data and are categorized as BCUs. Distinguishing between these two blazar subclasses is essential, as it provides insights into different physical emission mechanisms, jet properties, and facilitates population studies of blazars. To classify the BCUs, ML algorithms were trained on the $\gamma$-ray spectral and temporal properties of 2,219 labeled blazars (1,456 BL Lacs and 794 FSRQs) from the Fermi-LAT 4FGL catalog. The analysis employed three ML algorithms—Artificial Neural Networks (ANNs), XGBoost, and LightGBM. The dataset includes 18 input features derived $\gamma$-ray spectral properties (e.g., photon index and flux across various energy bands) and temporal characteristics (flux measurements collected over 12 years) for each source. The dataset was divided into training (80\%) and testing (20\%) subsets. A 15-fold cross-validation was applied, further enhancing the reliability of the results. In this process, the dataset is divided into 15 subsets, with the model trained on 14 subsets while the remaining subset is used for validation. This process repeats until each subset has served as the validation set, enabling a comprehensive evaluation of model performance across different data segments. To achieve optimal performance, hyperparameter tuning was used on the XGBoost and LightGBM models using Bayesian optimization via the HyperOpt package. This process identified key parameters that enhanced model performance, including a learning rate of 0.3, 400 iterations, and 31 leaf nodes. The higher learning rate and greater leaf count helped the model capture complex relationships within the data without overfitting, while the substantial number of iterations improved accuracy for this moderately sized dataset. Additionally, LightGBM and XGBoost are well-suited for tabular data, as they learn optimal branching directions during training and dynamically handle missing data, outperforming more complex models like ANNs, which are prone to overfitting with smaller datasets. With these parameters, the LightGBM model achieved a recall and precision of 88\%, making it the best-performing model for classifying BCUs. This model accurately classified 1,493 BCUs in the dataset, identifying 825 as BL Lac candidates and 405 as FSRQ candidates, while leaving 190 BCUs unclassified.
\begin{figure*}
    \centering
    \includegraphics[width=0.48\linewidth]{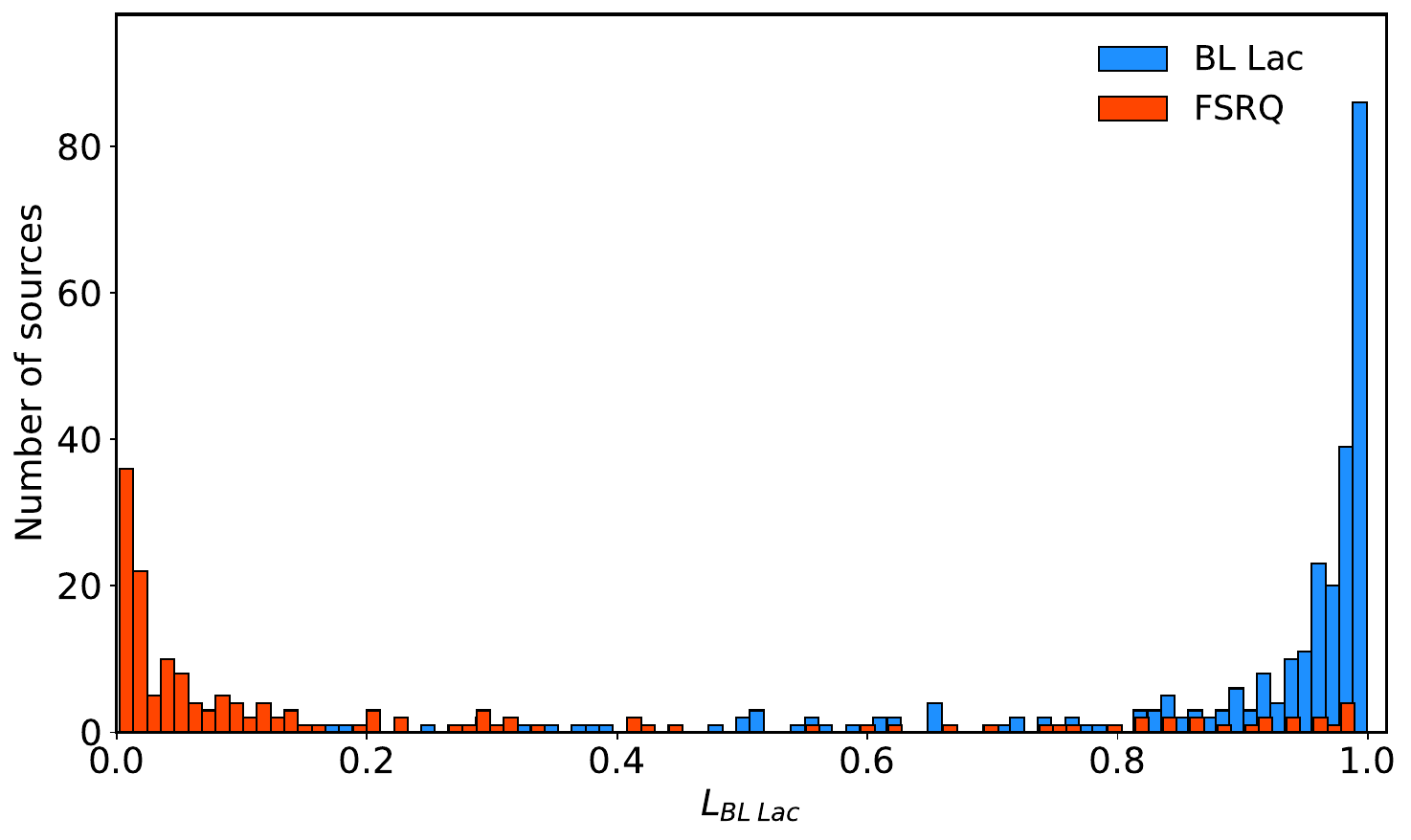}
    \includegraphics[width=0.48\linewidth]{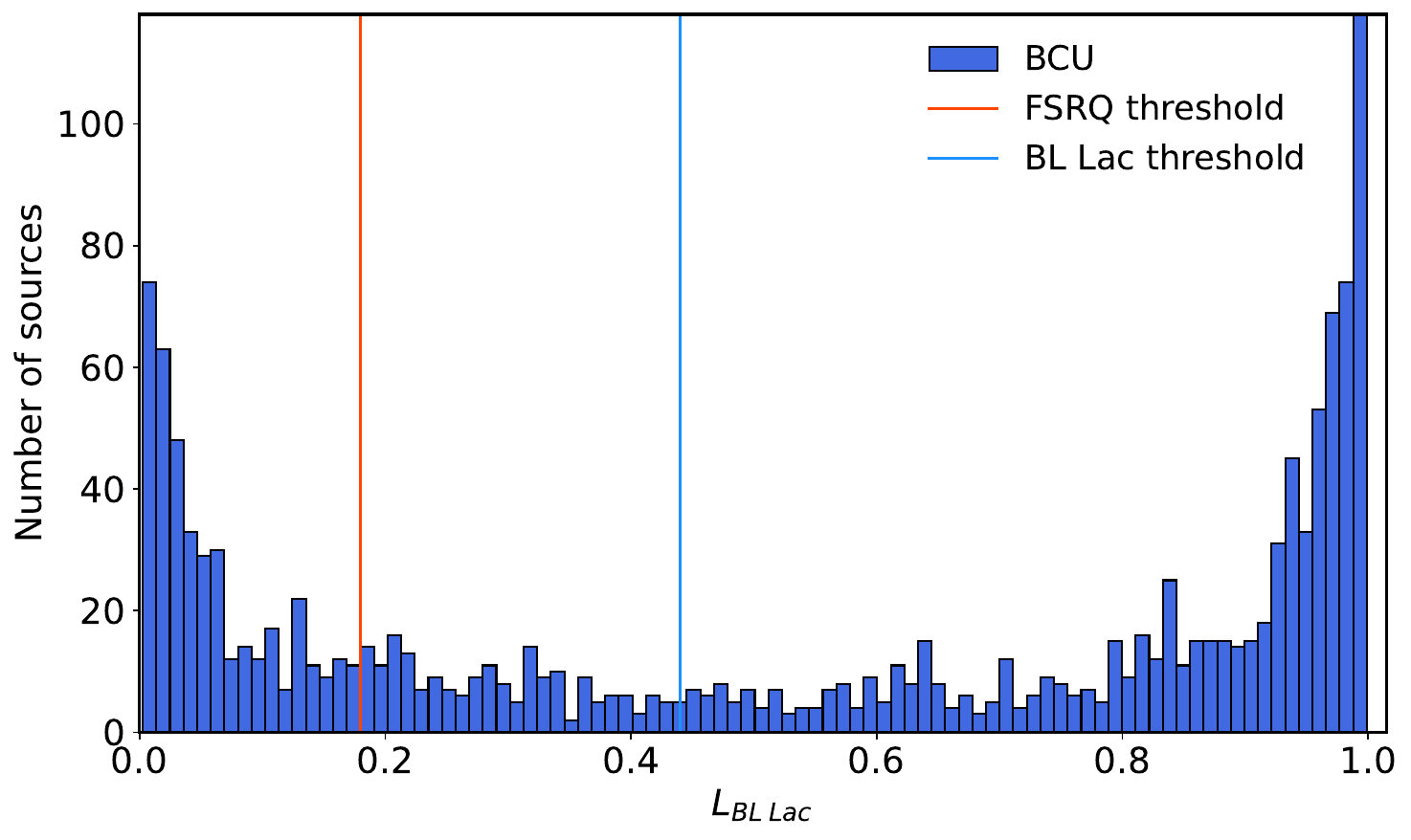}
    \caption{Likelihood distributions for sources in the test sample (left panel) and BCUs (right panel), showing probabilities of being classified as BL Lac or FSRQ. Adapted from Ref. ~[7].}

    \label{fig:bcu}
\end{figure*}

Fig. \ref{fig:bcu} displays the likelihood distribution produced by the optimized LightGBM model, demonstrating its effectiveness in distinguishing between BL Lacs and FSRQs. In the test sample (left panel), two distinct peaks are visible: BL Lacs (shown in blue) cluster toward higher likelihood values for BL Lac classification, while FSRQs (orange) cluster at lower values, corresponding to FSRQ classification likelihood. This clear separation indicates the accuracy of the model in classifying the test sample, thereby validating its performance. The right panel of Fig. \ref{fig:bcu} extends this analysis to 1,420 BCUs, showing a similar trend that straightness the robustness of the model  when applied to unclassified sources.

The classification results not only reduce the percentage of BCUs in the Fermi-LAT catalog but also provided a comprehensive view of the blazar population emitting in the $\gamma$-ray band. This approach enabled comparisons of various characteristics (such as $\gamma$-ray photon index, synchrotron and inverse Compton peaks) between BL Lacs and FSRQs, allowing for a quantitative study of their population distributions \cite{2023MNRAS.519.3000S}. These findings demonstrate the use of ML in identifying patterns within high-dimensional astrophysical data, which, in this case, contributes to a deeper understanding of blazar subclasses.
\section{Modeling blazar SEDs}
Fitting the spectral energy distributions (SEDs) of blazars with numerical models is a challenging task, largely due to the computational cost of such simulations. The radiative processes involved, including synchrotron self-Compton (SSC) and external inverse Compton (EIC) emissions, require intricate and resource-intensive calculations to produce accurate results. This complexity increases when models are considered in time-dependent contexts, as it requires solving kinetic equations that account for particle acceleration, cooling, and interactions within the jet environment—making conventional numerical fitting methods time-consuming. To overcome these computational challenges, a novel approach using neural networks, particularly CNN, has been developed in Refs.~\refcite{2024ApJ...963...71B} and \refcite{2024ApJ...971...70S}. This approach is designed to reduce computational time while preserving high accuracy in SED modeling. By training a CNN on a large set of simulated SEDs, this method bypasses the need for repeated numerical calculations, providing an efficient alternative for parameter exploration and real-time fitting of observed blazar spectra. In this framework, an extensive range of physically realistic parameters is explored for both SSC and EIC models to ensure comprehensive coverage of all relevant parameters. For the SSC model, 200,000 parameter combinations were generated using Latin hypercube sampling, a method that ensures uniform coverage of the parameter space while avoiding the regular spacing associated with grid-based approaches. For the EIC model, which includes additional parameters to account for external photon fields, 1 million parameter sets were generated. These parameter sets were then used as input into Simulator of Processes in Relativistic AstroNomical Objects (SOPRANO) code \cite{2022MNRAS.509.2102G} to produce the corresponding spectra. SOPRANO is designed for efficient, detailed computation of radiative signatures in relativistic astrophysical sources, using implicit numerical methods to solve the Fokker-Planck equation for particle distributions and the integro-differential equations governing photon distributions. This enables SOPRANO to capture a wide range of physical processes in generating realistic SEDs for each parameter configuration.
\begin{figure*}
    \centering
    \includegraphics[width=0.85\linewidth]{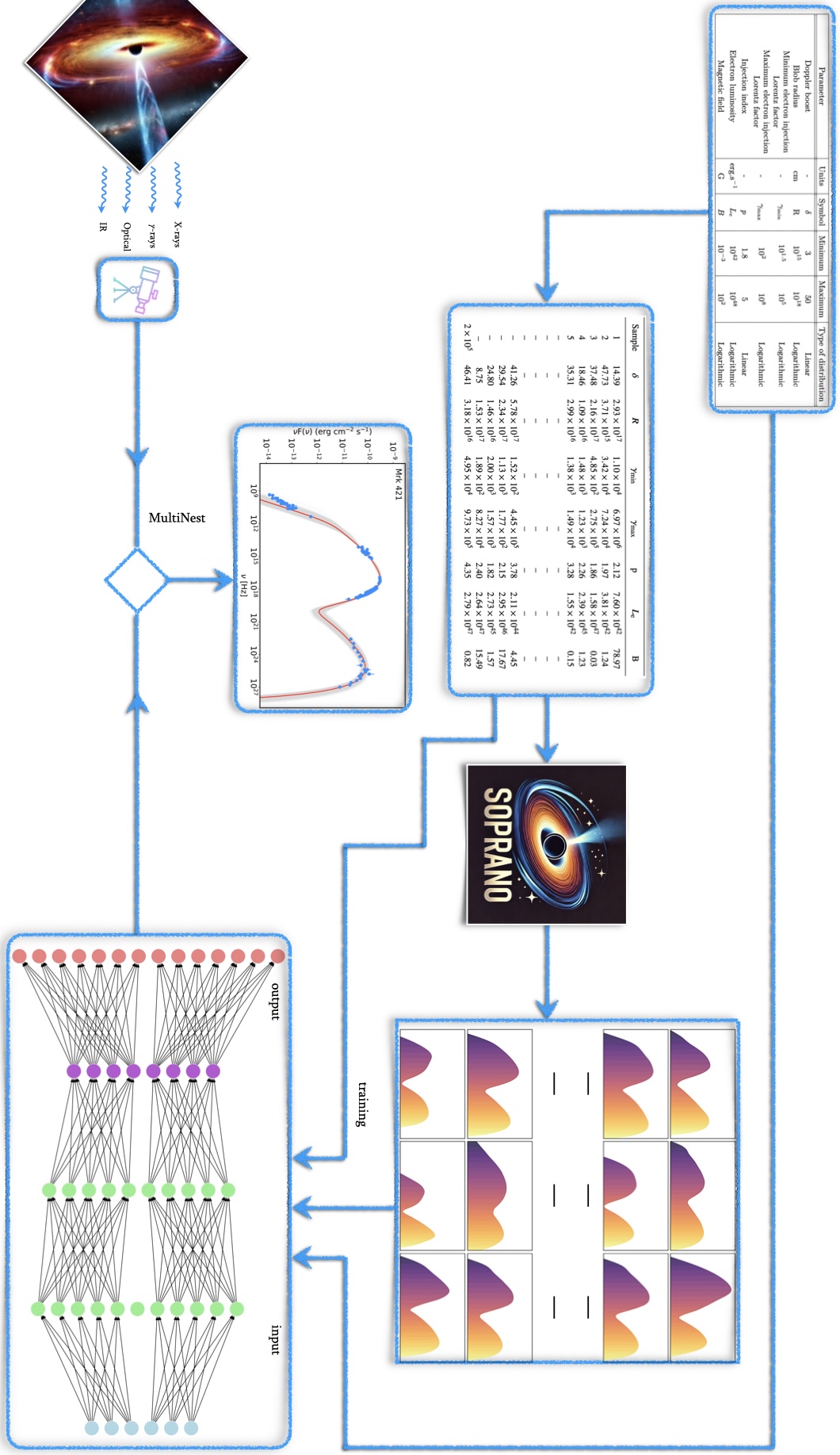}
    \caption{Workflow of the method for training a CNN for blazar SED modeling. A subset of parameters is selected from the entire parameter space using Latin hypercube sampling and passed to SOPRANO to compute the corresponding SEDs. Both the entire parameter range and the selected subset, along with the generated SEDs, are then passed to the CNN, which learns and predicts the relationships between all parameters and SEDs. This network combined to MultiNest can fit the observed data.}
    \label{fig:CNN}
\end{figure*}
The SEDs generated by SOPRANO serve as the training dataset for the CNN, which is optimized to capture the complex spectral features accurately. During training, the network learns the relationship between input parameters and their corresponding SEDs. Once trained, the CNN can predict SEDs across the full parameter range. In addition to data preparation and principal component analysis, further measures are implemented during training to prevent oscillations in the predicted spectra. Specifically, linear combinations of neighboring spectral points are constrained to limit independent fluctuations, which ensures smooth transitions and minimizes artifacts in the CNN-generated SEDs. The trained CNN model demonstrates excellent performance, accurately predicting SEDs across the parameter space while reducing computational time from several seconds or minutes to just milliseconds per evaluation. This efficiency enables the CNN to reproduce the detailed radiative signatures of blazar jets, making it a practical surrogate model in blazar SED modeling. Fig. \ref{fig:CNN} illustrates the main steps and methodology for this method.
\begin{figure*}
    \centering
    \includegraphics[width=0.48\linewidth]{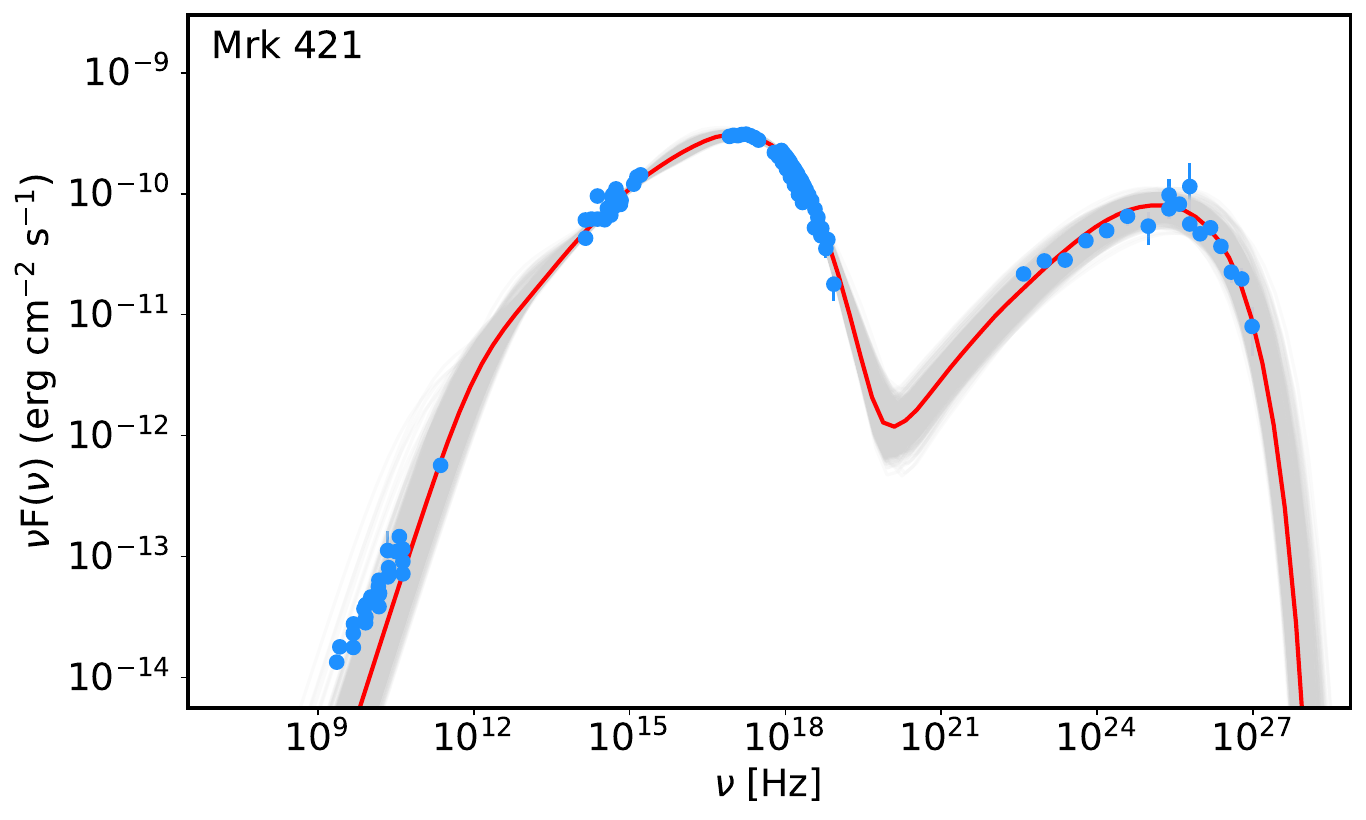}
    \includegraphics[width=0.48\linewidth]{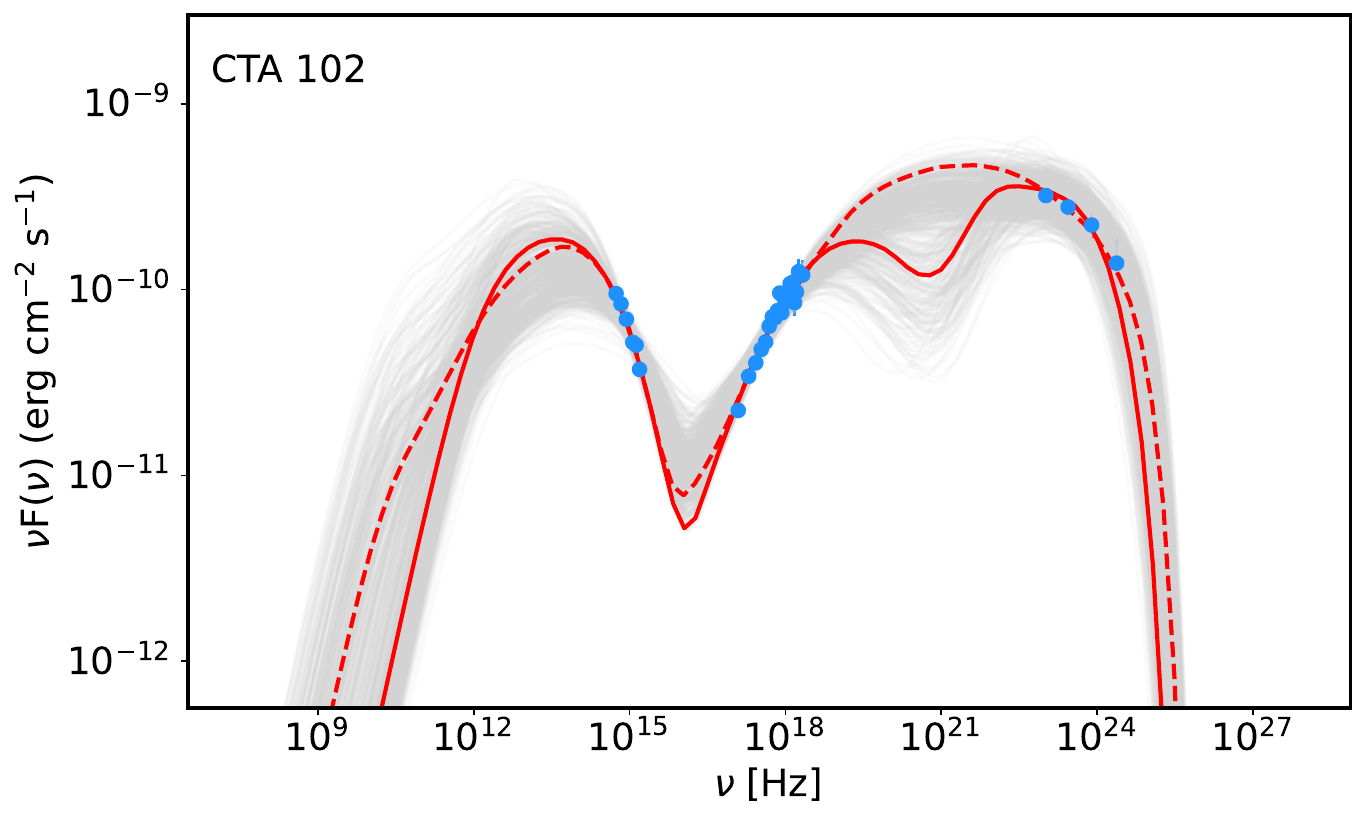}
    \caption{The broadband SEDs of Mrk 421 and CTA 102 are modeled under the SSC and EIC scenarios, respectively. The plots are adapted from Refs. 10 and 11.}
    \label{fig:sed}
\end{figure*}

Once trained, the CNN was applied to model the SEDs of specific blazars, such as Mrk 421 and CTA 102, as shown in Fig. \ref{fig:sed}. For Mrk 421 (Fig. \ref{fig:sed}, left), a SSC model was used to fit the observed broadband emission by coupling the trained CNN with the MultiNest \cite{2009MNRAS.398.1601F} algorithm for parameter optimization. Similarly, for CTA 102 (Fig. \ref{fig:sed}, right), the EIC model was used which includes also the interactions with external photon fields, showing the CNN ability to handle complex modeling scenarios. The CNN accurately models the SEDs for both sources, successfully capturing key spectral features and providing parameter estimates that align with theoretical expectations. These applications demonstrate the CNN capability to produce robust parameter constraints from observed SEDs across different blazar classes, enabling real-time analysis and interpretation of multiwavelength data. These CNNs are available for the researchers through Markarian Multiwavelength Data Center \cite{2024arXiv241001207S} (MMDC; www.mmdc.am).

This CNN-based framework represents a novel and groundbreaking approach to modeling blazar SEDs. Although it has been used here to model the broadband SEDs of blazars, the method is universal and can be adapted for various astrophysical sources, provided that a representative set of spectra can be computed for training. This approach presents a significant advancement in complex modeling of astrophysical sources by using neural networks to overcome computational limitations. The ability to rapidly and accurately reproduce radiative signatures across different astrophysical sources and environments has the potential to open new horizons in astrophysical research, enabling deeper and more efficient exploration of complex emission processes.

\section{Feature with Generative AI}
Generative AI represents a transformative advancement in ML, particularly through Large Language Models (LLMs), which are trained on vast datasets to process and generate human-like responses. These models operate on the basic principle of identifying patterns within their training data, enabling them to produce coherent and contextually relevant outputs based on user inputs. The versatility of generative AI in understanding and generating language, as well as other forms of data, has made it an invaluable tool across multiple fields. Since the introduction of ChatGPT in November 2022 \footnote{https://chatgpt.com}, generative AI has been widely used, rapidly becoming a useful tool in various sectors as well as integrating into daily life. This development has facilitated the automation of various tasks, such as customer service, content creation, language translation, personal assistance, and more. Its ability to manage, organize, and generate information efficiently has demonstrated its effectiveness in solving complex tasks, establishing it as a unique tool for supporting both professional and personal applications.

As generative AI successfully integrates into everyday tasks, it is clear that it will impact scientific research as well. The field of research functions as a collaborative ecosystem, where experienced scientists contribute knowledge and insights, guiding young researchers and students toward exploring specific, complex questions. This collaborative model relies heavily on effective communication, learning new knowledge, and continuously interpreting new data — in all these areas, generative AI can provide substantial support. As the volume of scientific literature and data continues to expand exponentially, researchers face challenges in staying up-to-date. For example, not regularly monitoring platforms such as arXiv \footnote{https://arxiv.org}, where new research papers are published daily, may result in missing critical developments. However, the limited time and ability of researchers to keep up with new information, combined with the ever-increasing number of research articles published each year, pose significant challenges. In contrast, the ongoing development of generative AI models, such as increasing the number of parameters used during training, enhances their versatility and improves the coherence and reasoning in the text they generate. Therefore, generative AI has the potential to serve as an invaluable assistant within the research ecosystem, supporting scientists, young researchers, and students in navigating, summarizing, and interpreting vast amounts of information. By processing complex data, generating hypotheses and proposing methodologies, generative AI can greatly enhance the productivity of the researchers, allowing them to focus on innovative problem-solving and deeper analysis.

It is essential to keep in mind the limitations of current generative AI models and to avoid the assumption that these models can easily fulfill all the needs of the researchers. First, while it is impressive how generative AI models process language and manage large datasets, they are not infallible. Second, these models may struggle with specialized scientific terminology, complex concepts, and sub domain-specific terminology, occasionally producing errors or “hallucinations” (i.e., generating possible but inaccurate information). Therefore, while generative AI has the potential to enhance research productivity, careful oversight by human experts, along with domain-specific adaptations and rigorous validation processes, is essential. This 'collaboration' between AI and human researchers has the potential to make research more dynamic, efficient, and innovative.

In recent years, various tools have been developed to integrate advances in generative AI into astrophysics, such as astroBERT \cite{2024ASPC..535..119G}, astroLLaMA\cite{2023arXiv230906126D}, pathfinder\cite{2024arXiv240801556I}, etc. The primary functionality of these tools is to summarize literature and conduct searches using databases created from the abstracts of scientific articles. This significantly improves accessibility to specific literature, making it possible to search for and find articles based on thematic keywords. However, the scope of these tools is largely limited to text-based applications. To overcome this limitation, we are developing a new tool, \textit{astroLLM}, which uses the full potential of generative AI for astrophysical research. This tool will not only contain deep knowledge in astrophysics but will also have access to multiwavelength datasets and the ability to perform theoretical modeling tasks. Its versatility surpasses that of current tools by integrating access to both data and modeling capabilities, making it an ideal research assistant for scientists. The main idea behind \textit{astroLLM} is to use the full potential of generative AI in astrophysics, ultimately increasing discovery potential. By combining data access, analysis, and modeling into a single platform, \textit{astroLLM} aims to streamline the research workflow and accelerate scientific advancements.
\section{Issues and Challenges}
The application of AI in research appears limitless, and it is very likely that in the near future, AI will completely transform scientific research. However, this change comes with significant challenges that the scientific community must address. These challenges are related to \textit{objectivity}, \textit{reproducibility}, \textit{transparency}, and \textit{accountability}, which, if not addressed, may lead to research misconduct.
To address these risks, the scientific community needs to take active steps which includes creating clear guidelines for using AI in research, making AI algorithms and datasets open and transparent for checking, encouraging thorough peer review, and emphasizing accountability with a focus on ethics. By addressing these challenges, we can fully benefit from the potential of AI while maintaining high scientific standards.

Currently, the main arguments against the application of AI in research, are biases, errors, and the "black box" problem. Although ML/AI tools are designed to accurately represent the data they are trained on, they can sometimes introduce biases, such as favoring specific patterns or answers. While bias in research can be addressed, the bias arising from AI—due to the scale and complexity of data, algorithms, and applications—can be difficult to control. It is the responsibility of the scientists that are using AI in research to identify, describe, and control bias which involves taking care of data diversity, sampling, and representativeness when training AI models. Bias, like errors, can significantly affect the validity and reliability of results. In the context of AI, error refers to the difference between the AI output and the correct output. While humans also make errors, the use of AI can substantially increase both the rate and magnitude of errors, potentially impacting research outcomes. Scientists using AI should  identify, describe, reduce, and correct potential errors (e.g., by discussing limitations and known AI-related issues), ensuring that only random errors, which are unavoidable, remain.

Both bias and error in AI are manageable challenges; if AI is used in a context where scientists expertise and judgment are sufficient to identify and address these issues, both can be effectively managed. However, the "black box" problem poses significant challenges to the trustworthiness and transparency of AI applications in research. This problem arises from the lack of complete understanding of AI models, making it difficult to fully trust and rely on their outputs. The concern is rooted in the principle that the effective application of any tool or technology depends on understanding how it works. For example, during the development of a telescope, considerable time is devoted to the commissioning phase to fully understand its operations and functions; otherwise, the data it gathers would be unreliable. However, there is no straightforward solution to the "black box" problem in AI. Even for experts, it is often impossible to delve deeply into the system, inspect the code, review input layers and weights, or trace the billions of steps executed by the network. A key implication of this challenge is the necessity to make both the data and full details of the AI system (e.g., model type, architecture, technical specifications, and optimization methods) accessible and transparent. Providing this information as supplementary material when using AI in research ensures reproducibility and significantly enhances the accountability and trustworthiness of AI-driven findings.

Current AI/ML methods are trained on large datasets for specific applications and, in certain tasks, can match or even surpass human experts by processing vast amounts of data efficiently. However, these methods have limitations and challenges, as outlined above, that pose significant difficulties for the smooth application of AI in various fields, including research. Addressing these challenges and limitations of current AI/ML models is essential in the near future, unless advancements lead to the development of Artificial General Intelligence (AGI), also known as human-level AI. AGIis a theoretical form of AI that matches human intelligence, allowing it to understand complex aspects, learn independently, apply knowledge across diverse fields, and adapt to new situations. Realizing AGI would revolutionize society, but it requires AI capabilities far exceeding those of current generative or task-specific models. Despite the rapid advancements in AI, AGI remains an ambitious and still unrealized objective. For now, researchers can choose to either proceed without incorporating AI (particularly generative AI), preserving creativity and reasoning as uniquely human endeavors, or take measured steps to integrate AI responsibly into research. By carefully using AI powerful tools  while keeping in mind its limitations, we can dramatically increase the potential for discovery in science. AI should be applied in contexts where scientists can fully understand and interpret the data and outputs, ensuring careful validation and trust in the results. This requires that, before applying AI algorithms to data, researchers \textit{(i)} thoroughly understand their datasets and \textit{(ii)} guide AI objectives while allowing it the flexibility to identify novel patterns that could lead to unexpected insights. Such a balanced approach enables the full potential of AI/ML tools to be realized, positioning them as valuable assistants in research that accelerate progress and expand discovery opportunities. In research, we should move beyond the debate of biological versus artificial intelligence —which often centers on the differences between human cognition and machine processing— and instead adopt \textbf{Human Guided Artificial Intelligence (HG-AI)} - AI systems directed by human intelligence. HG-AI emphasizes the synergy between human expertise and machine efficiency, using the strengths of both to achieve greater outcomes. By integrating human intuition and contextual understanding with the data-processing capabilities of AI, HG-AI can solve complex problems more effectively than either could alone.
By actively guiding AI to produce specific, interpretable outcomes, scientists can ensure that AI serves as a valuable collaborator, enhancing research while maintaining human oversight and understanding. This collaborative approach enables science to fully harness the advantages of AI while preserving the essential role of human creativity and reasoning.

\section{Conclusion}
The integration of AI in research, particularly in astrophysics, marks a transformative shift in data analysis, modeling, and prediction capabilities. In astrophysics, ML tools, ranging from supervised classification models to complex neural networks, have demonstrated their ability to handle and analyze the massive datasets generated by modern telescopes. Applications of these tools, such as the classification of diverse astrophysical sources and the modeling of processes like multiwavelength emissions from blazars, highlight the role of AI in improving the efficiency and depth of astrophysical research.

However, the deployment of AI also poses challenges, particularly regarding transparency, bias, and the “black box” nature of advanced models. The HG-AI framework offers a solution to these challenges: by emphasizing human oversight as a core element, it enables researchers to harness the strengths of AI while ensuring that results remain scientifically robust and interpretable. This approach advocates for responsible AI applications, where scientists actively guide and validate AI outputs to achieve rigorous and unbiased results. Such synergy is essential and holds the potential to accelerate discoveries in research.

\section*{Acknowledgments}

This work was support by the Higher Education and Science Committee of the Republic of Armenia, in the frames of the research project No 23LCG-1C004.



\end{document}